\documentclass[%
prl,
twocolumn,
groupedaddress,
showkeys,
amsmath,
amssymb,
]{revtex4-1}
\usepackage{graphicx}
\usepackage{dcolumn}
\usepackage{bm}
\usepackage{caption}
\usepackage{subfloat}
\usepackage{subfig}
\usepackage{float}

\begin{document}


\title{Trivial and Non-trivial Superconductivity in dsDNA}


\author{H. Simchi}
\email{simchi@alumni.iust.ac.ir}
\affiliation{Department of Physics, Iran University of Science and Technology, Narmak, Tehran 16844, Iran}
\affiliation{Semiconductor Technoloy Center, P.O. Box 19575-199, Tehran, Iran} 


\date{\today}

\begin{abstract}
A double-stranded DNA (dsDNA) is modeled by two coupled one-dimensional Kitaev's chain and the topological superconductivity is studied. It is shown that the zero energy mode exists under some specific conditions. The wave function of zero mode is calculated and it is shown that the Majorana quasi-particles exist on the ends of each strand. By calculating the winding number, we show that the topological phase transition can happen if the hopping integral between two strands is very smaller than the pairing potential between the Cooper pairs. It means that the dsDNA behaves as a trivial superconductor, commonly, but single-stranded DNA  (or two coupled ssDNA with very small hopping between them) may behave as a non-trivial superconductor. Finally, we suggest an experimental setup for probable detection of Majorana quasi-particle in DNA.
\end{abstract}


\maketitle

Certain materials can conduct electric current without any loss, which are called superconductors. It means that, If we consider a superconducting loop with initial current $I_{0\ }$at time $t_0$, and the magnetic field surrounding the loop is measured to be same after a time$\ t_1$, say one year later, it appears to mean that the $I\left(t_1\right)=I\left(t_0\right)=I_0$ \cite{R1}.By decreasing the temperature of the superconductive materials below a critical temperature, $T_c$, the current carriers are paired. The paired carriers, called Cooper pairs, move inside the superconductors with no resistance, and in consequence, the superconductivity phenomenon is seen in these materials \cite{R1}. In a spinless p-wave superconductor and in strong pairing regime, a molecule-like Cooper pairs form from two fermions bound in the real space, whereas in the weak paring regime the Cooper pair size is infinite \cite{R2}. The probable phase transition between strong and weak regimes is the main subject of the topological superconductivity (TSC) field \cite{R2,R3}. The topological materials are classified into ten symmetry classes called Altland-Zirnbauer (AZ) symmetry classes \cite{R3,R4}. Of course, the ten symmetry classes can be simplified as an eightfold periodic classification \cite{R3}. It should be noted that the weak and strong regimes are always separated by a quantum phase transition if the symmetry conditions are strictly enforced \cite{R3}.This implies that if these two regimes are in spatial proximity, there should be a gapless state (called Majorana bound state (MBS)) localized at the boundary between the two phases. The presence of such gapless boundary states is the main feature of TSC and can be considered as a definition for these kind of materials \cite{R3}.

\noindent        The requirements for making the Majorana fermions are so generic: Take a superconductor, remove degeneracy by breaking spin-rotation and time-reversal symmetries, then close and reopen the excitation gap \cite{R5}. The main pathways for creating the MBS are \cite{R5}: Shockley mechanism \cite{R6}, Topological insulators \cite{R7}, Semiconductor heterostructures \cite{R8}, and chiral p-wave superconductors \cite{R6, R9, Rten, R11}. Many theoretical and experimental works have been done for studying the Majorana creation and detection, before in Refs. \cite{R12, R13,R14,R15,R16,R17,R18,R19,R20,R21,R22,R23,R24}.

\noindent         It is shown that the Deoxyribonucleic acid (DNA) could be an insulator in room temperature, a semiconductor with a wide band gap in all temperatures, a resistance in room temperature, a metal in low temperature and a superconductor at $T<1K$ \cite{R25}. For studying the proximity-induced superconductivity in DNA, a rhenium carbon (Re/C) electrode was deposited on a mica substrate by sputtering method. Then, the $\lambda$DNA molecule was deposited on the Re/C electrode and the temperature was set to $<1K$ in Ref. \cite{R26}. For explaining the observed superconductivity effect in DNA, Ren et al. showed that the hopping integral between bases changes due to the temperature fluctuation of twist angle between them and consequently, the electric conductivity increases \cite{R27}. We have used the Bogoliubov-de-Gennes (BdG) equation and shown that the pairing potential is equal to 0.76$\mu$eV \cite{R28}, and the perfect spin-polarization can happen in dsDNA \cite{R29}.  

\noindent         In this letter, a double-stranded DNA (dsDNA) is modeled by two coupled one-dimensional Kitaev's chain\cite{R30}, and the topological superconductivity is studied. The eigenfunctions and eigenvalues are calculated, and it is shown that the zero energy mode exists under some specific conditions. The wave function of zero mode is calculated in coordinate-space, and it is shown that the MBS exist on the ends of each strand. We will show that the topological phase transition can happen if the hopping integral between two strands is very smaller than the pairing potential between the Cooper pairs in dsDNA. It means that the dsDNA behaves as a trivial superconductor, commonly (as it has been seen, before in Ref. \cite{R26}). However, single-stranded DNA (or two coupled ssDNA without hopping between them) may behave as a non-trivial superconductor. Finally, we suggest an experimental set up for probable detection of Majorana in DNA.

The ladder model was introduced for studying the charge transport in DNA in Ref. \cite{R31}. In this model, the electrons hope between Adenine and Thymine, with hopping integral $t_1$, on the first strand, and they also hope between Guanine and Cytosine, with hopping integral $t_2$, on the second strand. The hopping between two strands is possible, too \cite{R29,R31, R32, R33}. Also, it was shown that for a specific range of electron energies and for $t_1=t_2=t$ the perfect spin-polarization could be seen due to the spin-orbit coupling (SOC) effect in dsDNA in Ref. \cite{R29}. Therefore, by mixing the ladder model and Kitaev's model, we can write the Hamiltonian of the dsDNA in momentum-space as:
\begin{equation} \label{GrindEQ__1_} 
H=\left({\epsilon }_k{\sigma }_z+\widetilde{\Delta }{\sigma }_y\right)I_{\tau }+\lambda I_{\sigma }{\tau }_x  ,
\end{equation} 
where ${\epsilon }_k=-tcosk-\mu $ and $\widetilde{\Delta }=-\Delta sink$ \cite{R2,R30}. In fact, Eq. (1) is Fourier transform of a ladder model (on each strand) which is mixed with Cooper pairing mechanism. Here, we assume that the lattice constant is equal to one for simplicity. $t$ is hopping integral between two neighborhood sites on each strand, $\lambda $ is hopping integral between two strands and $\mu $ is the chemical potential. ${\sigma }_i,\ i=x,y,z$ is general Pauli matrix and ${\tau }_i,\ i=x,y,z$ is Pauli matrix in strand space. $I_{\tau }$ ($I_{\sigma }$) is unit matrix in strand (general) space. By simple calculations, it can be shown that the eigenvalues are:
\begin{equation} \label{GrindEQ__2_} 
E_i={\left(-1\right)}^n\lambda +{\left(-1\right)}^m\sqrt{{{\epsilon }_k}^2+{\widetilde{\Delta }}^2}    , i=1,2,3,4 ,
\end{equation} 
where $n=2\eqref{GrindEQ__1_}$ for $i=1,2\ (3,4)$ and $m=2\eqref{GrindEQ__1_}$ for $i=1,4\ (2,3)$ .                                                        

\noindent    By using the equation $\psi =E\psi $, we can find the eigenfunctions which are
\begin{widetext}
\begin{equation}  \label{GrindEQ__3_}
{\psi }_i={\left(\frac{{\epsilon }_k+{\left(-1\right)}^{i+1}\sqrt{{{\epsilon }_k}^2+{\widetilde{\Delta }}^2}}{-i\widetilde{\Delta }},\ 1,\frac{{\epsilon }_k+{\left(-1\right)}^{i+1}\sqrt{{{\epsilon }_k}^2+{\widetilde{\Delta }}^2}}{-i\widetilde{\Delta }},1\right)}^T,                
\end{equation}
\end{widetext}
for $i=1,2$  and ${E=E}_i$ and 
\begin{widetext}
\begin{equation}  \label{GrindEQ__4_}
{\psi }_i={\left(\frac{{-\epsilon }_k+{\left(-1\right)}^i\sqrt{{{\epsilon }_k}^2+{\widetilde{\Delta }}^2}}{-i\widetilde{\Delta }},\ -1,\frac{{\epsilon }_k+{\left(-1\right)}^{i+1}\sqrt{{{\epsilon }_k}^2+{\widetilde{\Delta }}^2}}{-i\widetilde{\Delta }},1\right)}^T,                
\end{equation}
\end{widetext} 
for $i=3,$4 and ${E=E}_i$ .
\noindent However for small $\widetilde{\Delta }$ (means large $\mu $) the energy eigenvalues are equal to
\begin{equation} \label{GrindEQ__5_} 
E_i={\left(-1\right)}^n\lambda +{\left(-1\right)}^m{\epsilon }_k    , i=1,2,3,4 .                                                              
\end{equation} 
Now, the Fermi level is set by ${\epsilon }_k=\mu =0$ ($\mu =0$ means half filling) \cite{R34}. This gives the two Fermi points $k_F=\pm \pi /2$ . If $k=k_F$ then $E_i={\left(-1\right)}^n\lambda +{\left(-1\right)}^{m+1}\mu $ and the energy gap,  $\Delta E\neq 0$. By increasing $\widetilde{\Delta }$ (means decreasing $\mu $), if $k=k_F$, then $E_i={\left(-1\right)}^n\lambda +{\left(-1\right)}^m\sqrt{{\mu }^2+{\Delta }^2}$. Now for $\lambda =\pm \sqrt{{\mu }^2+{\Delta }^2}$ the gap is closed ($\Delta E=0$). Now, if $\widetilde{\Delta }$ increases more (means $\mu $ decrease more) the gap $\Delta E\neq 0$ and increases (see Fig. 1). Therefore, if we want to have a phase transition from low pairing situation (i.e., small $\widetilde{\Delta }$) to high pairing situation (i.e., large $\widetilde{\Delta }$) the gap closure should happen. It means that we encounter a topological phase transition which can be tuned by adjusting the chemical potential $\mu $ through a gate voltage (as an example). A question can be asked: what about the behavior of wavefunction when the energy gap is closed?

\begin{figure}
\includegraphics[width=\columnwidth]{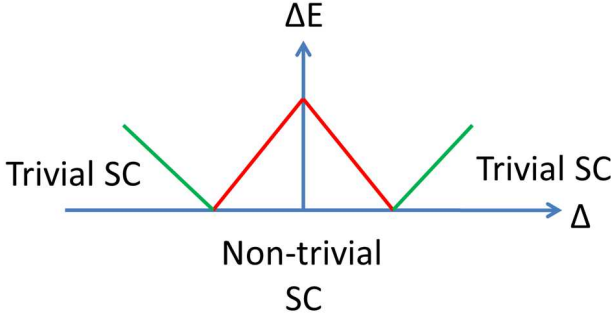}
\caption{\label{fig:epsart} (Color online) General Schematic of energy gap variations via pairing potential.}
\end{figure}

\noindent         We can expand $cosk$ and $sink$ around Fermi points and write the Eq. \eqref{GrindEQ__1_} as below
\begin{equation} \label{GrindEQ__6_} 
H=\left(-it{\partial }_x{\sigma }_z+\Delta (x){\sigma }_y\right)I_{\tau }+\lambda I_{\sigma }{\tau }_x  ,                                                                        
\end{equation} 
where $k=-i{\partial }_x$. The wave function can be written as $\psi ={\left({\phi }_1\left(x\right),{\phi }_2\left(x\right),{\phi }_3\left(x\right),{\phi }_4\left(x\right)\right)}^T$. If  ${\phi }_1={\phi }_2={\varphi }_1$ and ${\phi }_3={\phi }_4={\varphi }_2$ (see Eqs. (3) and (4)) and we define $F={\varphi }_1{\varphi }_2$, it can be easily shown that for zero energy case $E=0$:
\begin{equation} \label{GrindEQ__7_} 
{\partial }_xF+\left(\frac{\Delta (x)}{t}\right)F=0  .                                                                                                         
\end{equation} 
Therefore, 
\begin{equation} \label{GrindEQ__8_} 
F\left(x\right)={F\left(0\right)\ e}^{-\int{\frac{\Delta \left(x\right)}{t}}\ dx}  .                                                                                              
\end{equation} 
Now if $\Delta (x)$ varies slowly in space as: $\Delta \left(x\right)={\Delta }_0tanh\left(x/\xi \right)$, where $\xi $ is a large length scale characterizing the slowly varying gap function ($\xi \gg $ lattice constant) \cite{R34} then
\begin{equation} \label{GrindEQ__9_} 
F={\varphi }_1{\varphi }_2=F(0){\left(\mathrm{cosh}\mathrm{}(\frac{x}{\xi })\right)}^{-{\mathrm{\Delta }}_0\xi /t} .
\end{equation} 
Now since ${\varphi }_2=F/{\varphi }_1$, it can be shown that
\begin{equation} \label{GrindEQ__10_} 
t{\partial }_x\left({{\varphi }_1}^2\right)+i\lambda {{\varphi }_1}^2=\mathrm{\Delta }F .
\end{equation} 
Therefore,
\begin{equation} \label{GrindEQ__11_} 
{{\varphi }_1}^2=e^{-i\lambda x/t}\int{\frac{-\mathrm{\Delta }F(x)}{t}}e^{i\lambda x/t}\ dx ,
\end{equation} 
and
\begin{equation} \label{GrindEQ__12_} 
{{\varphi }_2}^2=e^{i\lambda x/t}\int{\frac{-\mathrm{\Delta }F(x)}{t}}e^{-i\lambda x/t}\ dx .
\end{equation} 
Thus, there are the zero energy Fermion states.  For $x\to \pm \infty $ and $\lambda \to 0$ (although ${-1\le e}^{\pm i\lambda x/t}\le +1$) then $\int{\frac{-\mathrm{\Delta }F(x)}{t}}e^{\pm i\lambda x/t}\ dx\to \int{\frac{-\mathrm{\Delta }F(x)}{t}}\ dx$ . Now, by using Eq. \eqref{GrindEQ__9_} we find
\begin{equation} \label{GrindEQ__13_} 
\int{\frac{-\mathrm{\Delta }F(x)}{t}}\ dx=-{\left(cosh\left(x/\xi \right)\right)}^{\frac{-{\Delta }_0\xi }{t}}\to e^{-\left|x\right|/\alpha } ,
\end{equation} 
where, $\alpha =t/{\Delta }_0$. Therefore, the zero energy Fermion states are localized at the ends of each strand. However, the condition $\lambda \to 0$ means that we have two parallel ssDNA. As a consequence, the ssDNA (or two coupled ssDNA with very small hopping between them) may behave as a topological superconductor and the zero energy Fermion states are localized at its ends. In next, we will also see that the condition $\lambda \to 0$ causes the non-vanishing winding number.                   

\noindent             Of course, for $\widetilde{\Delta }={\Delta }_0=$cte, it can be shown that ${\varphi }_1=\sqrt{\frac{-{\Delta }_0}{i\lambda -2{\mathrm{\Delta }}_0}}e^{\frac{-{\mathrm{\Delta }}_0\left|x\right|}{2t}}$  and  ${\varphi }_2=\sqrt{\frac{{\Delta }_0}{i\lambda +2{\mathrm{\Delta }}_0}}e^{\frac{-{\mathrm{\Delta }}_0\left|x\right|}{2t}}$. Since they decay exponentially as $x\to \pm \infty $ the zero energy Fermion states are localized at the end of each strand of dsDNA, again.

\noindent         For calculating the winding number, we can use the below formula \cite{R21}
\begin{equation} \label{GrindEQ__14_} 
v=\frac{1}{4\pi }\int^{k_2}_{k_1}{dk\ Tr\left(CH^{-1}\partial_{k}H\right)} ,
\end{equation} 
where $C$ is particle-hole symmetry operator. By using Eq. \eqref{GrindEQ__1_}, it can be shown that
\begin{equation} \label{GrindEQ__15_} 
H^{-1}=\left(a{\sigma }_z+b{\sigma }_y\right)I_{\tau }+cI_{\sigma }{\tau }_x ,
\end{equation} 
where $a={\epsilon }_k/\left({{\epsilon }_k}^2+{\widetilde{\mathrm{\Delta }}}^2-{\lambda }^2\right)$,
$b=\widetilde{\mathrm{\Delta }}/\left({{\epsilon }_k}^2+{\widetilde{\mathrm{\Delta }}}^2-{\lambda }^2\right)$, and
$c=-\lambda /\left({{\epsilon }_k}^2+{\widetilde{\mathrm{\Delta }}}^2-{\lambda }^2\right)$.

\noindent Since $C={\sigma }_xI_{\tau }$ then
\begin{equation} \label{GrindEQ__16_} 
Tr\left(CH^{-1}\partial_{k}H\right)=4\left(b{\partial }_k{\epsilon }_k-a{\partial }_k\widetilde{\Delta }\right) ,
\end{equation}
or
\begin{equation} \label{GrindEQ__16_}
Tr\left(CH^{-1}H\right)=\frac{4}{\left({{\epsilon }_k}^2+{\widetilde{\mathrm{\Delta }}}^2-{\lambda }^2\right)}\left(\widetilde{\Delta }{\partial }_k{\epsilon }_k-{\epsilon }_k{\partial }_k\widetilde{\Delta }\right) .
\end{equation} 
If we use polar coordinates and define $cos\theta =\frac{{\epsilon }_k}{\sqrt{\left({{\epsilon }_k}^2+{\widetilde{\mathrm{\Delta }}}^2\right)}}$  and  $sin\theta =\frac{\widetilde{\Delta }}{\sqrt{\left({{\epsilon }_k}^2+{\widetilde{\mathrm{\Delta }}}^2\right)}}$ then
\begin{equation} \label{GrindEQ__17_} 
d\theta =\frac{1}{\left({{\epsilon }_k}^2+{\widetilde{\mathrm{\Delta }}}^2\right)}\left({\epsilon }_k{\partial }_k\widetilde{\Delta }-\widetilde{\Delta }{\partial }_k{\epsilon }_k\right)\ dk .
\end{equation} 
Therefore
\begin{equation} \label{GrindEQ__18_} 
v=\frac{1}{4\pi }\int^{k_2}_{k_1}{dk}Tr\left(CH^{-1}H\right)=\frac{1}{\pi }\int^{{\theta }_2}_{{\theta }_1}{\{d\theta }\frac{\left({{\epsilon }_k}^2+{\widetilde{\mathrm{\Delta }}}^2\right)}{\left({{\epsilon }_k}^2+{\widetilde{\mathrm{\Delta }}}^2-{\lambda }^2\right)}\} 
\end{equation} 
Now if ${\widetilde{\mathrm{\Delta }}}^2\gg {\lambda }^2$ (or$\ \lambda \to 0$), then  $v=2\int^{{\theta }_2}_{{\theta }_1}{\frac{d\theta }{2\pi }}$  where $\int^{{\theta }_2}_{{\theta }_1}{\frac{d\theta }{2\pi }}$  is winding number of a single strand (i.e., a Kitaev's chain).  It means when pairing potential is very greater than the hopping integral between two strands, each ssDNA behaves as a non-trivial superconductor.   
It should be noted that for $\widetilde{\Delta }=\lambda$, it can be shown that $v=(tan\theta)^{\theta_{2}}_{\theta_{1}}$. Thus, as one sweep $k$ from zero to $\pi$, $\theta_{1}$ is equal to zero and $\theta_{2}=0$ or $\pi$. However, for $\theta_{1}$ both values of $\theta_{2}$, the winding number will be equal to zero, and as a consequence, dsDNA is a trivial superconductor.

The Kitaev's model is spinless \cite{R30}. Therefore, one question can be asked: how is the assumed dsDNA model spinless?  It was shown that the spin splitting happens in dsDNA when the SOC term is added to the Hamiltonian in Refs. \cite{R32,R33}. We have shown that the perfect spin polarization can be seen in dsDNA for a range of hopping integral on each strand and electron energy (using BdG equation) \cite{R29}. In addition, it has been shown that the spin-singlet Cooper pairing cannot induce proximity superconductivity in a perfect spin-polarized system, but the p-wave pairing is nonzero only due to the SOC and weakness as the Zeeman field increases and the spins become increasingly polarized \cite{R35}. This implies that the SOC quantum wire realizes a topological superconductivity phase in the limit of strong Zeeman field \cite{R35}. By using first-order perturbation theory, it has been shown that the SOC term only modifies pairing potential as $\Delta \to \frac{{\lambda }_{so}\mathrm{\Delta }}{B}$ in BdG equation where ${\lambda }_{so}$ and $B$ are SOC and magnetic field strength, respectively \cite{R35}. Therefore, we should add SOC and magnetic field to Eq. (1) for finding the perfect spin polarization in dsDNA \cite{R29,R32,R33}. However, instead of doing that, we can only change the pairing potential in the above calculations \cite{R35} and choose a suitable set of parameters in Eq. (1) by using the results of Ref. 29. 

\noindent      Finally, we try to introduce an experimental setup for detecting the zero energy modes (i.e., MBS) as follows. Gohler et al., have measured the spin polarization in a self-assembled monolayer of dsDNA \cite{R36}. They have deposited four different lengths of dsDNA on a clean gold substrate. The photoelectrons were ejected by an ultraviolet laser pulse, and the kinetic energy distribution of electrons and spin analysis were found by using an electron time-of-flight instrument and a Mott-type electron polarimeter, respectively \cite{R36}. Devillard et al., have used a superconductivity tunnel microscope for finding the fingerprints of Majorana fermions in topological nanowires (TN) \cite{R23}. The superconductor tip (SCT) is connected to a spin filter (T-junction connected to two polarized quantum dots) for studying the spin current correlations between SCT and the TN. Using a properly oriented spin filter, the spin current correlations are always negative for an MBS \cite{R23}. Kasumov et al. have deposited DNA on Re/C substrate and observed the proximity-induced superconductivity effect in DNA \cite{R26}. Based on these done experimental works, we can imagine a setup including a suitable substrate such that some type I superconductors (TISC) (e.g Re/C) are deposited on its surface by the sputtering method (see Fig. 2).  Then, we should deposit a monolayer of dsDNA on the Re/C electrodes and find the suitable conditions for finding perfect spin polarization i.e., we should find a suitable TISC and a suitable set of experimental parameters. Now we can decrease the temperature below the critical temperature of the TISC and adjust the normal metal tip of a tunneling microscope at the ends of the strands of DNA. By measuring the zero bias conductance we may find the MBS. We hope our suggested setup and obtained results can motivate experimentalists for searching Majorana fermions in DNA.

\begin{figure}
\includegraphics[width=\columnwidth]{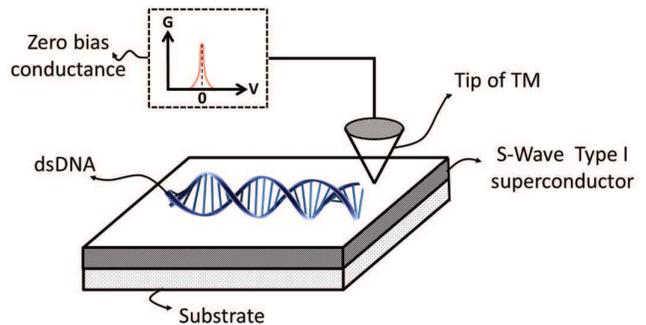}
\caption{\label{fig:epsart} (Color online) General schematic of an experimental setup for detecting Majorana quasi-particle in DNA.}
\end{figure}

 In parallel with other attempts for finding Majorana fermions using parallel nanowires \cite{R37}, based on the published experimental results about proximity-induced superconductivity \cite{R26} and spin-polarization in DNA \cite{R36}, and existing measuring methods for finding the effect of Majorana bound states on zero bias conductance \cite{R23}, we have mixed the ladder and Kitaev's model and shown that the dsDNA is commonly a trivial superconductor but two attached ssDNA (i.e., dsDNA with very small hopping between two strands) or a ssDNA can be a non-trivial superconductor. Moreover, we have proposed an experimental setup generally for checking the explained theory in the letter. We hope that the results can motivate experimentalists for testing the probable topological superconductivity in DNA.

\nocite{*}
\bibliography{article}

\providecommand{\noopsort}[1]{}\providecommand{\singleletter}[1]{#1}%
\begin{thebibliography}{37}%
\makeatletter
\providecommand \@ifxundefined [1]{%
 \@ifx{#1\undefined}
}%
\providecommand \@ifnum [1]{%
 \ifnum #1\expandafter \@firstoftwo
 \else \expandafter \@secondoftwo
 \fi
}%
\providecommand \@ifx [1]{%
 \ifx #1\expandafter \@firstoftwo
 \else \expandafter \@secondoftwo
 \fi
}%
\providecommand \natexlab [1]{#1}%
\providecommand \enquote  [1]{``#1''}%
\providecommand \bibnamefont  [1]{#1}%
\providecommand \bibfnamefont [1]{#1}%
\providecommand \citenamefont [1]{#1}%
\providecommand \href@noop [0]{\@secondoftwo}%
\providecommand \href [0]{\begingroup \@sanitize@url \@href}%
\providecommand \@href[1]{\@@startlink{#1}\@@href}%
\providecommand \@@href[1]{\endgroup#1\@@endlink}%
\providecommand \@sanitize@url [0]{\catcode `\\12\catcode `\$12\catcode
  `\&12\catcode `\#12\catcode `\^12\catcode `\_12\catcode `\%12\relax}%
\providecommand \@@startlink[1]{}%
\providecommand \@@endlink[0]{}%
\providecommand \url  [0]{\begingroup\@sanitize@url \@url }%
\providecommand \@url [1]{\endgroup\@href {#1}{\urlprefix }}%
\providecommand \urlprefix  [0]{URL }%
\providecommand \Eprint [0]{\href }%
\providecommand \doibase [0]{http://dx.doi.org/}%
\providecommand \selectlanguage [0]{\@gobble}%
\providecommand \bibinfo  [0]{\@secondoftwo}%
\providecommand \bibfield  [0]{\@secondoftwo}%
\providecommand \translation [1]{[#1]}%
\providecommand \BibitemOpen [0]{}%
\providecommand \bibitemStop [0]{}%
\providecommand \bibitemNoStop [0]{.\EOS\space}%
\providecommand \EOS [0]{\spacefactor3000\relax}%
\providecommand \BibitemShut  [1]{\csname bibitem#1\endcsname}%
\let\auto@bib@innerbib\@empty
\bibitem [{\citenamefont {Fossheim}\ and\ \citenamefont {Sudbo}(2004)}]{R1}%
  \BibitemOpen
  \bibfield  {author} {\bibinfo {author} {\bibfnamefont {K.}~\bibnamefont
  {Fossheim}}\ and\ \bibinfo {author} {\bibfnamefont {A.}~\bibnamefont
  {Sudbo}},\ }\href@noop {} {\emph {\bibinfo {title} {Superconductivity:
  Physics and Applications}}}\ (\bibinfo  {publisher} {John Wiley and Sons
  Ltd},\ \bibinfo {year} {2004})\BibitemShut {NoStop}%
\bibitem [{\citenamefont {Alicea}(2012)}]{R2}%
  \BibitemOpen
  \bibfield  {author} {\bibinfo {author} {\bibfnamefont {J.}~\bibnamefont
  {Alicea}},\ }\href@noop {} {\bibfield  {journal} {\bibinfo  {journal} {Rep.
  Prog. Phys.}\ }\textbf {\bibinfo {volume} {75}},\ \bibinfo {pages} {076501}
  (\bibinfo {year} {2012})}\BibitemShut {NoStop}%
\bibitem [{\citenamefont {Chiu}\ \emph {et~al.}(2016)\citenamefont {Chiu},
  \citenamefont {Teo}, \citenamefont {Schnyder},\ and\ \citenamefont
  {Ryu}}]{R3}%
  \BibitemOpen
  \bibfield  {author} {\bibinfo {author} {\bibfnamefont {C.~K.}\ \bibnamefont
  {Chiu}}, \bibinfo {author} {\bibfnamefont {J.~C.~Y.}\ \bibnamefont {Teo}},
  \bibinfo {author} {\bibfnamefont {A.~P.}\ \bibnamefont {Schnyder}}, \ and\
  \bibinfo {author} {\bibfnamefont {S.}~\bibnamefont {Ryu}},\ }\href@noop {}
  {\bibfield  {journal} {\bibinfo  {journal} {, Rev. Mod. Phys.}\ }\textbf
  {\bibinfo {volume} {88}},\ \bibinfo {pages} {035005} (\bibinfo {year}
  {2016})}\BibitemShut {NoStop}%
\bibitem [{\citenamefont {Benalcazar}\ \emph {et~al.}(2014)\citenamefont
  {Benalcazar}, \citenamefont {Teo},\ and\ \citenamefont {Hughes}}]{R4}%
  \BibitemOpen
  \bibfield  {author} {\bibinfo {author} {\bibfnamefont {W.~A.}\ \bibnamefont
  {Benalcazar}}, \bibinfo {author} {\bibfnamefont {J.~C.~Y.}\ \bibnamefont
  {Teo}}, \ and\ \bibinfo {author} {\bibfnamefont {T.~L.}\ \bibnamefont
  {Hughes}},\ }\href@noop {} {\bibfield  {journal} {\bibinfo  {journal} {Phys.
  Rev. B}\ }\textbf {\bibinfo {volume} {89}},\ \bibinfo {pages} {224503}
  (\bibinfo {year} {2014})}\BibitemShut {NoStop}%
\bibitem [{\citenamefont {Beenakker}(2013)}]{R5}%
  \BibitemOpen
  \bibfield  {author} {\bibinfo {author} {\bibfnamefont {C.~W.}\ \bibnamefont
  {Beenakker}},\ }\href@noop {} {\bibfield  {journal} {\bibinfo  {journal}
  {Annu. Rev. Con. Mat. Phys.}\ }\textbf {\bibinfo {volume} {4}},\ \bibinfo
  {pages} {113} (\bibinfo {year} {2013})}\BibitemShut {NoStop}%
\bibitem [{\citenamefont {Wimmer}\ \emph {et~al.}(2010)\citenamefont {Wimmer},
  \citenamefont {Akhmerov}, \citenamefont {Medvedyeva}, \citenamefont
  {Tworzydlo},\ and\ \citenamefont {Beenakker}}]{R6}%
  \BibitemOpen
  \bibfield  {author} {\bibinfo {author} {\bibfnamefont {M.}~\bibnamefont
  {Wimmer}}, \bibinfo {author} {\bibfnamefont {A.~R.}\ \bibnamefont
  {Akhmerov}}, \bibinfo {author} {\bibfnamefont {M.~V.}\ \bibnamefont
  {Medvedyeva}}, \bibinfo {author} {\bibfnamefont {J.}~\bibnamefont
  {Tworzydlo}}, \ and\ \bibinfo {author} {\bibfnamefont {C.~W.~J.}\
  \bibnamefont {Beenakker}},\ }\href@noop {} {\bibfield  {journal} {\bibinfo
  {journal} {Phys. Rev. Lett.}\ }\textbf {\bibinfo {volume} {105}},\ \bibinfo
  {pages} {046803} (\bibinfo {year} {2010})}\BibitemShut {NoStop}%
\bibitem [{\citenamefont {Qi}\ and\ \citenamefont {Zhang}(2011)}]{R7}%
  \BibitemOpen
  \bibfield  {author} {\bibinfo {author} {\bibfnamefont {X.~L.}\ \bibnamefont
  {Qi}}\ and\ \bibinfo {author} {\bibfnamefont {S.~C.}\ \bibnamefont {Zhang}},\
  }\href@noop {} {\bibfield  {journal} {\bibinfo  {journal} {Rev. Mod. Phys.}\
  }\textbf {\bibinfo {volume} {83}},\ \bibinfo {pages} {1057} (\bibinfo {year}
  {2011})}\BibitemShut {NoStop}%
\bibitem [{\citenamefont {Alicea}(2010)}]{R8}%
  \BibitemOpen
  \bibfield  {author} {\bibinfo {author} {\bibfnamefont {J.}~\bibnamefont
  {Alicea}},\ }\href@noop {} {\bibfield  {journal} {\bibinfo  {journal} {Phys.
  Rev. B}\ }\textbf {\bibinfo {volume} {81}},\ \bibinfo {pages} {125318}
  (\bibinfo {year} {2010})}\BibitemShut {NoStop}%
\bibitem [{\citenamefont {Potter}\ and\ \citenamefont {Lee}(2010)}]{R9}%
  \BibitemOpen
  \bibfield  {author} {\bibinfo {author} {\bibfnamefont {A.}~\bibnamefont
  {Potter}}\ and\ \bibinfo {author} {\bibfnamefont {P.~A.}\ \bibnamefont
  {Lee}},\ }\href@noop {} {\bibfield  {journal} {\bibinfo  {journal} {Phys.
  Rev. Lett.}\ }\textbf {\bibinfo {volume} {105}},\ \bibinfo {pages} {227003}
  (\bibinfo {year} {2010})}\BibitemShut {NoStop}%
\bibitem [{\citenamefont {Lutchyn}\ \emph {et~al.}(2011)\citenamefont
  {Lutchyn}, \citenamefont {Stanescu},\ and\ \citenamefont {Sarma}}]{Rten}%
  \BibitemOpen
  \bibfield  {author} {\bibinfo {author} {\bibfnamefont {R.~M.}\ \bibnamefont
  {Lutchyn}}, \bibinfo {author} {\bibfnamefont {T.~D.}\ \bibnamefont
  {Stanescu}}, \ and\ \bibinfo {author} {\bibfnamefont {S.~D.}\ \bibnamefont
  {Sarma}},\ }\href@noop {} {\bibfield  {journal} {\bibinfo  {journal} {Phys.
  Rev. Lett.}\ }\textbf {\bibinfo {volume} {106}},\ \bibinfo {pages} {127001}
  (\bibinfo {year} {2011})}\BibitemShut {NoStop}%
\bibitem [{\citenamefont {Zhou}\ and\ \citenamefont {Shen}(2011)}]{R11}%
  \BibitemOpen
  \bibfield  {author} {\bibinfo {author} {\bibfnamefont {B.}~\bibnamefont
  {Zhou}}\ and\ \bibinfo {author} {\bibfnamefont {S.~Q.}\ \bibnamefont
  {Shen}},\ }\href@noop {} {\bibfield  {journal} {\bibinfo  {journal} {Phys.
  Rev. B}\ }\textbf {\bibinfo {volume} {84}},\ \bibinfo {pages} {054532}
  (\bibinfo {year} {2011})}\BibitemShut {NoStop}%
\bibitem [{\citenamefont {Bishara}\ \emph {et~al.}(2009)\citenamefont
  {Bishara}, \citenamefont {Bonderson}, \citenamefont {Nayak}, \citenamefont
  {Shtengel},\ and\ \citenamefont {Slingerland}}]{R12}%
  \BibitemOpen
  \bibfield  {author} {\bibinfo {author} {\bibfnamefont {W.}~\bibnamefont
  {Bishara}}, \bibinfo {author} {\bibfnamefont {P.}~\bibnamefont {Bonderson}},
  \bibinfo {author} {\bibfnamefont {C.}~\bibnamefont {Nayak}}, \bibinfo
  {author} {\bibfnamefont {K.}~\bibnamefont {Shtengel}}, \ and\ \bibinfo
  {author} {\bibfnamefont {K.}~\bibnamefont {Slingerland}},\ }\href@noop {}
  {\bibfield  {journal} {\bibinfo  {journal} {Phys. Rev. B }\ }\textbf
  {\bibinfo {volume} {80}},\ \bibinfo {pages} {155303} (\bibinfo {year}
  {2009})}\BibitemShut {NoStop}%
\bibitem [{\citenamefont {Domingnez}\ \emph {et~al.}(2012)\citenamefont
  {Domingnez}, \citenamefont {Hassler},\ and\ \citenamefont {Platero}}]{R13}%
  \BibitemOpen
  \bibfield  {author} {\bibinfo {author} {\bibfnamefont {F.}~\bibnamefont
  {Domingnez}}, \bibinfo {author} {\bibfnamefont {F.}~\bibnamefont {Hassler}},
  \ and\ \bibinfo {author} {\bibfnamefont {G.}~\bibnamefont {Platero}},\
  }\href@noop {} {\bibfield  {journal} {\bibinfo  {journal} {Phys. Rev. B }\
  }\textbf {\bibinfo {volume} {86}},\ \bibinfo {pages} {140503} (\bibinfo
  {year} {2012})}\BibitemShut {NoStop}%
\bibitem [{\citenamefont {Xu}\ \emph {et~al.}(2015)\citenamefont {Xu},
  \citenamefont {Wang}, \citenamefont {Liu}, \citenamefont {Ge}, \citenamefont
  {Yang}, \citenamefont {Liu}, \citenamefont {Xu}, \citenamefont {Guan},
  \citenamefont {Gao}, \citenamefont {Qian}, \citenamefont {Liu}, \citenamefont
  {Wang}, \citenamefont {F.~C.~Zhang},\ and\ \citenamefont {Jia}}]{R14}%
  \BibitemOpen
  \bibfield  {author} {\bibinfo {author} {\bibfnamefont {J.~P.}\ \bibnamefont
  {Xu}}, \bibinfo {author} {\bibfnamefont {M.~X.}\ \bibnamefont {Wang}},
  \bibinfo {author} {\bibfnamefont {Z.~L.}\ \bibnamefont {Liu}}, \bibinfo
  {author} {\bibfnamefont {J.~F.}\ \bibnamefont {Ge}}, \bibinfo {author}
  {\bibfnamefont {X.}~\bibnamefont {Yang}}, \bibinfo {author} {\bibfnamefont
  {C.}~\bibnamefont {Liu}}, \bibinfo {author} {\bibfnamefont {Z.~A.}\
  \bibnamefont {Xu}}, \bibinfo {author} {\bibfnamefont {D.}~\bibnamefont
  {Guan}}, \bibinfo {author} {\bibfnamefont {C.~L.}\ \bibnamefont {Gao}},
  \bibinfo {author} {\bibfnamefont {D.}~\bibnamefont {Qian}}, \bibinfo {author}
  {\bibfnamefont {Y.}~\bibnamefont {Liu}}, \bibinfo {author} {\bibfnamefont
  {Q.~H.}\ \bibnamefont {Wang}}, \bibinfo {author} {\bibfnamefont {Q.~K.~X.}\
  \bibnamefont {F.~C.~Zhang}}, \ and\ \bibinfo {author} {\bibfnamefont {J.~F.}\
  \bibnamefont {Jia}},\ }\href@noop {} {\bibfield  {journal} {\bibinfo
  {journal} {Phys. Rev. Lett.}\ }\textbf {\bibinfo {volume} {114}},\ \bibinfo
  {pages} {017001} (\bibinfo {year} {2015})}\BibitemShut {NoStop}%
\bibitem [{\citenamefont {Joes}\ \emph {et~al.}(2015)\citenamefont {Joes},
  \citenamefont {Lado}, \citenamefont {Aguado}, \citenamefont {Guinea},\ and\
  \citenamefont {Rossier}}]{R15}%
  \BibitemOpen
  \bibfield  {author} {\bibinfo {author} {\bibfnamefont {P.~S.}\ \bibnamefont
  {Joes}}, \bibinfo {author} {\bibfnamefont {J.~L.}\ \bibnamefont {Lado}},
  \bibinfo {author} {\bibfnamefont {R.}~\bibnamefont {Aguado}}, \bibinfo
  {author} {\bibfnamefont {F.}~\bibnamefont {Guinea}}, \ and\ \bibinfo {author}
  {\bibfnamefont {J.~F.}\ \bibnamefont {Rossier}},\ }\href@noop {} {\bibfield
  {journal} {\bibinfo  {journal} {Phys. Rev. X}\ }\textbf {\bibinfo {volume}
  {5}},\ \bibinfo {pages} {041042} (\bibinfo {year} {2015})}\BibitemShut
  {NoStop}%
\bibitem [{\citenamefont {Wang}\ and\ \citenamefont {Wu}(2016)}]{R16}%
  \BibitemOpen
  \bibfield  {author} {\bibinfo {author} {\bibfnamefont {L.}~\bibnamefont
  {Wang}}\ and\ \bibinfo {author} {\bibfnamefont {M.~W.}\ \bibnamefont {Wu}},\
  }\href@noop {} {\bibfield  {journal} {\bibinfo  {journal} {Phys. Rev. B}\
  }\textbf {\bibinfo {volume} {93}},\ \bibinfo {pages} {054502} (\bibinfo
  {year} {2016})}\BibitemShut {NoStop}%
\bibitem [{\citenamefont {Lv}\ \emph {et~al.}(2017)\citenamefont {Lv},
  \citenamefont {Wang}, \citenamefont {Zhang}, \citenamefont {Ding},
  \citenamefont {Li}, \citenamefont {Wang}, \citenamefont {He}, \citenamefont
  {Song}, \citenamefont {Ma},\ and\ \citenamefont {Xue}}]{R17}%
  \BibitemOpen
  \bibfield  {author} {\bibinfo {author} {\bibfnamefont {Y.~F.}\ \bibnamefont
  {Lv}}, \bibinfo {author} {\bibfnamefont {W.~L.}\ \bibnamefont {Wang}},
  \bibinfo {author} {\bibfnamefont {Y.~M.}\ \bibnamefont {Zhang}}, \bibinfo
  {author} {\bibfnamefont {H.}~\bibnamefont {Ding}}, \bibinfo {author}
  {\bibfnamefont {W.}~\bibnamefont {Li}}, \bibinfo {author} {\bibfnamefont
  {L.}~\bibnamefont {Wang}}, \bibinfo {author} {\bibfnamefont {K.}~\bibnamefont
  {He}}, \bibinfo {author} {\bibfnamefont {C.~L.}\ \bibnamefont {Song}},
  \bibinfo {author} {\bibfnamefont {X.~C.}\ \bibnamefont {Ma}}, \ and\ \bibinfo
  {author} {\bibfnamefont {Q.~K.}\ \bibnamefont {Xue}},\ }\href@noop {}
  {\bibfield  {journal} {\bibinfo  {journal} {Science Bulletin }\ }\textbf
  {\bibinfo {volume} {62}},\ \bibinfo {pages} {852} (\bibinfo {year}
  {2017})}\BibitemShut {NoStop}%
\bibitem [{\citenamefont {Sun}\ and\ \citenamefont {Jia}(2017)}]{R18}%
  \BibitemOpen
  \bibfield  {author} {\bibinfo {author} {\bibfnamefont {H.~H.}\ \bibnamefont
  {Sun}}\ and\ \bibinfo {author} {\bibfnamefont {J.~F.}\ \bibnamefont {Jia}},\
  }\href@noop {} {\bibfield  {journal} {\bibinfo  {journal} {Natu. Part. J.
  Quant. Matter}\ }\textbf {\bibinfo {volume} {2}},\ \bibinfo {pages} {34}
  (\bibinfo {year} {2017})}\BibitemShut {NoStop}%
\bibitem [{\citenamefont {Dutreix}(2017)}]{R19}%
  \BibitemOpen
  \bibfield  {author} {\bibinfo {author} {\bibfnamefont {C.}~\bibnamefont
  {Dutreix}},\ }\href@noop {} {\bibfield  {journal} {\bibinfo  {journal} {Phys.
  Rev. B}\ }\textbf {\bibinfo {volume} {96}},\ \bibinfo {pages} {045416}
  (\bibinfo {year} {2017})}\BibitemShut {NoStop}%
\bibitem [{\citenamefont {Hsu}\ \emph {et~al.}(2017)\citenamefont {Hsu},
  \citenamefont {Vaezi}, \citenamefont {Fischer},\ and\ \citenamefont
  {Kim}}]{R20}%
  \BibitemOpen
  \bibfield  {author} {\bibinfo {author} {\bibfnamefont {Y.~T.}\ \bibnamefont
  {Hsu}}, \bibinfo {author} {\bibfnamefont {A.}~\bibnamefont {Vaezi}}, \bibinfo
  {author} {\bibfnamefont {M.~H.}\ \bibnamefont {Fischer}}, \ and\ \bibinfo
  {author} {\bibfnamefont {E.~A.}\ \bibnamefont {Kim}},\ }\href@noop {}
  {\bibfield  {journal} {\bibinfo  {journal} {Natu. Comm.}\ }\textbf {\bibinfo
  {volume} {8}},\ \bibinfo {pages} {14985} (\bibinfo {year}
  {2017})}\BibitemShut {NoStop}%
\bibitem [{\citenamefont {Sahlberg}\ \emph {et~al.}(2017)\citenamefont
  {Sahlberg}, \citenamefont {Weststrom}, \citenamefont {Poyhonen},\ and\
  \citenamefont {Ojanen}}]{R21}%
  \BibitemOpen
  \bibfield  {author} {\bibinfo {author} {\bibfnamefont {.~I.}\ \bibnamefont
  {Sahlberg}}, \bibinfo {author} {\bibfnamefont {A.}~\bibnamefont {Weststrom}},
  \bibinfo {author} {\bibfnamefont {K.}~\bibnamefont {Poyhonen}}, \ and\
  \bibinfo {author} {\bibfnamefont {T.}~\bibnamefont {Ojanen}},\ }\href@noop {}
  {\bibfield  {journal} {\bibinfo  {journal} {Phys. Rev. B }\ }\textbf
  {\bibinfo {volume} {95}},\ \bibinfo {pages} {184512} (\bibinfo {year}
  {2017})}\BibitemShut {NoStop}%
\bibitem [{\citenamefont {Suominen}\ \emph {et~al.}(2017)\citenamefont
  {Suominen}, \citenamefont {Kjaergaard}, \citenamefont {Hamilton},
  \citenamefont {Shabani}, \citenamefont {Palmstrom}, \citenamefont {Marcus},\
  and\ \citenamefont {Nichele}}]{R22}%
  \BibitemOpen
  \bibfield  {author} {\bibinfo {author} {\bibfnamefont {H.~J.}\ \bibnamefont
  {Suominen}}, \bibinfo {author} {\bibfnamefont {M.}~\bibnamefont
  {Kjaergaard}}, \bibinfo {author} {\bibfnamefont {A.~R.}\ \bibnamefont
  {Hamilton}}, \bibinfo {author} {\bibfnamefont {J.}~\bibnamefont {Shabani}},
  \bibinfo {author} {\bibfnamefont {C.~J.}\ \bibnamefont {Palmstrom}}, \bibinfo
  {author} {\bibfnamefont {C.~M.}\ \bibnamefont {Marcus}}, \ and\ \bibinfo
  {author} {\bibfnamefont {F.}~\bibnamefont {Nichele}},\ }\href@noop {}
  {\bibfield  {journal} {\bibinfo  {journal} {Phys. Rev. Lett.}\ }\textbf
  {\bibinfo {volume} {119}},\ \bibinfo {pages} {176805} (\bibinfo {year}
  {2017})}\BibitemShut {NoStop}%
\bibitem [{\citenamefont {Devillard}\ \emph {et~al.}(2017)\citenamefont
  {Devillard}, \citenamefont {Chevallier},\ and\ \citenamefont {Albert}}]{R23}%
  \BibitemOpen
  \bibfield  {author} {\bibinfo {author} {\bibfnamefont {P.}~\bibnamefont
  {Devillard}}, \bibinfo {author} {\bibfnamefont {D.}~\bibnamefont
  {Chevallier}}, \ and\ \bibinfo {author} {\bibfnamefont {M.}~\bibnamefont
  {Albert}},\ }\href@noop {} {\bibfield  {journal} {\bibinfo  {journal} {Phys.
  Rev. B}\ }\textbf {\bibinfo {volume} {96}},\ \bibinfo {pages} {115413}
  (\bibinfo {year} {2017})}\BibitemShut {NoStop}%
\bibitem [{\citenamefont {Zhang}\ \emph {et~al.}(2017)\citenamefont {Zhang},
  \citenamefont {Gul}, \citenamefont {Boj}, \citenamefont {Nowak},
  \citenamefont {Wimmer}, \citenamefont {Zuo}, \citenamefont {Mourik},
  \citenamefont {de~Vries}, \citenamefont {van Veen}, \citenamefont {de~Moor},
  \citenamefont {Bommer}, \citenamefont {van Woerkim}, \citenamefont {Car},
  \citenamefont {Plissard}, \citenamefont {Bakkers}, \citenamefont {Perez},
  \citenamefont {Cassidy}, \citenamefont {Koelling}, \citenamefont {Goswami},
  \citenamefont {Watanabe}, \citenamefont {Taniguchi},\ and\ \citenamefont
  {Kouwenhoven}}]{R24}%
  \BibitemOpen
  \bibfield  {author} {\bibinfo {author} {\bibfnamefont {H.}~\bibnamefont
  {Zhang}}, \bibinfo {author} {\bibfnamefont {O.}~\bibnamefont {Gul}}, \bibinfo
  {author} {\bibfnamefont {S.~C.}\ \bibnamefont {Boj}}, \bibinfo {author}
  {\bibfnamefont {M.~P.}\ \bibnamefont {Nowak}}, \bibinfo {author}
  {\bibfnamefont {M.}~\bibnamefont {Wimmer}}, \bibinfo {author} {\bibfnamefont
  {K.}~\bibnamefont {Zuo}}, \bibinfo {author} {\bibfnamefont {V.}~\bibnamefont
  {Mourik}}, \bibinfo {author} {\bibfnamefont {F.~K.}\ \bibnamefont
  {de~Vries}}, \bibinfo {author} {\bibfnamefont {J.}~\bibnamefont {van Veen}},
  \bibinfo {author} {\bibfnamefont {M.~W.~A.}\ \bibnamefont {de~Moor}},
  \bibinfo {author} {\bibfnamefont {J.~D.~S.}\ \bibnamefont {Bommer}}, \bibinfo
  {author} {\bibfnamefont {D.~J.}\ \bibnamefont {van Woerkim}}, \bibinfo
  {author} {\bibfnamefont {D.}~\bibnamefont {Car}}, \bibinfo {author}
  {\bibfnamefont {S.~R.}\ \bibnamefont {Plissard}}, \bibinfo {author}
  {\bibfnamefont {E.~P. A.~M.}\ \bibnamefont {Bakkers}}, \bibinfo {author}
  {\bibfnamefont {M.~Q.}\ \bibnamefont {Perez}}, \bibinfo {author}
  {\bibfnamefont {M.~C.}\ \bibnamefont {Cassidy}}, \bibinfo {author}
  {\bibfnamefont {S.}~\bibnamefont {Koelling}}, \bibinfo {author}
  {\bibfnamefont {S.}~\bibnamefont {Goswami}}, \bibinfo {author} {\bibfnamefont
  {K.}~\bibnamefont {Watanabe}}, \bibinfo {author} {\bibfnamefont
  {T.}~\bibnamefont {Taniguchi}}, \ and\ \bibinfo {author} {\bibfnamefont
  {L.~P.}\ \bibnamefont {Kouwenhoven}},\ }\href@noop {} {\bibfield  {journal}
  {\bibinfo  {journal} {Nature Comm.}\ }\textbf {\bibinfo {volume} {8}},\
  \bibinfo {pages} {16025} (\bibinfo {year} {2017})}\BibitemShut {NoStop}%
\bibitem [{\citenamefont {Endres}\ \emph {et~al.}(2004)\citenamefont {Endres},
  \citenamefont {Cox},\ and\ \citenamefont {Singh}}]{R25}%
  \BibitemOpen
  \bibfield  {author} {\bibinfo {author} {\bibfnamefont {R.~G.}\ \bibnamefont
  {Endres}}, \bibinfo {author} {\bibfnamefont {D.~L.}\ \bibnamefont {Cox}}, \
  and\ \bibinfo {author} {\bibfnamefont {R.~R.~P.}\ \bibnamefont {Singh}},\
  }\href@noop {} {\bibfield  {journal} {\bibinfo  {journal} {Rev. Mode. Phys.}\
  }\textbf {\bibinfo {volume} {76}},\ \bibinfo {pages} {195} (\bibinfo {year}
  {2004})}\BibitemShut {NoStop}%
\bibitem [{\citenamefont {Kasumov}\ \emph {et~al.}(2001)\citenamefont
  {Kasumov}, \citenamefont {Kociak}, \citenamefont {Gueron}, \citenamefont
  {Rulet}, \citenamefont {Volkov}, \citenamefont {Klinov},\ and\ \citenamefont
  {Bouchiat}}]{R26}%
  \BibitemOpen
  \bibfield  {author} {\bibinfo {author} {\bibfnamefont {A.~Y.}\ \bibnamefont
  {Kasumov}}, \bibinfo {author} {\bibfnamefont {M.}~\bibnamefont {Kociak}},
  \bibinfo {author} {\bibfnamefont {S.}~\bibnamefont {Gueron}}, \bibinfo
  {author} {\bibfnamefont {B.}~\bibnamefont {Rulet}}, \bibinfo {author}
  {\bibfnamefont {V.~T.}\ \bibnamefont {Volkov}}, \bibinfo {author}
  {\bibfnamefont {D.~V.}\ \bibnamefont {Klinov}}, \ and\ \bibinfo {author}
  {\bibfnamefont {H.}~\bibnamefont {Bouchiat}},\ }\href@noop {} {\bibfield
  {journal} {\bibinfo  {journal} {Science}\ }\textbf {\bibinfo {volume}
  {292}},\ \bibinfo {pages} {280} (\bibinfo {year} {2001})}\BibitemShut
  {NoStop}%
\bibitem [{\citenamefont {Ren}\ \emph {et~al.}(2005)\citenamefont {Ren},
  \citenamefont {Wang}, \citenamefont {Ma},\ and\ \citenamefont {Guo}}]{R27}%
  \BibitemOpen
  \bibfield  {author} {\bibinfo {author} {\bibfnamefont {W.}~\bibnamefont
  {Ren}}, \bibinfo {author} {\bibfnamefont {J.}~\bibnamefont {Wang}}, \bibinfo
  {author} {\bibfnamefont {Z.}~\bibnamefont {Ma}}, \ and\ \bibinfo {author}
  {\bibfnamefont {H.}~\bibnamefont {Guo}},\ }\href@noop {} {\bibfield
  {journal} {\bibinfo  {journal} {Phys. Rev. B}\ }\textbf {\bibinfo {volume}
  {72}},\ \bibinfo {pages} {035456} (\bibinfo {year} {2005})}\BibitemShut
  {NoStop}%
\bibitem [{\citenamefont {Simchi}\ \emph {et~al.}(2014)\citenamefont {Simchi},
  \citenamefont {Esmaeilzadeh},\ and\ \citenamefont {Mazidabadi}}]{R28}%
  \BibitemOpen
  \bibfield  {author} {\bibinfo {author} {\bibfnamefont {H.}~\bibnamefont
  {Simchi}}, \bibinfo {author} {\bibfnamefont {M.}~\bibnamefont
  {Esmaeilzadeh}}, \ and\ \bibinfo {author} {\bibfnamefont {H.}~\bibnamefont
  {Mazidabadi}},\ }\href@noop {} {\bibfield  {journal} {\bibinfo  {journal} {J.
  Appl. Phys.}\ }\textbf {\bibinfo {volume} {115}},\ \bibinfo {pages} {054702}
  (\bibinfo {year} {2014})}\BibitemShut {NoStop}%
\bibitem [{\citenamefont {Simchi}\ \emph {et~al.}(2013)\citenamefont {Simchi},
  \citenamefont {Esmaeilzadeh},\ and\ \citenamefont {Mazidabadi}}]{R29}%
  \BibitemOpen
  \bibfield  {author} {\bibinfo {author} {\bibfnamefont {H.}~\bibnamefont
  {Simchi}}, \bibinfo {author} {\bibfnamefont {M.}~\bibnamefont
  {Esmaeilzadeh}}, \ and\ \bibinfo {author} {\bibfnamefont {H.}~\bibnamefont
  {Mazidabadi}},\ }\href@noop {} {\bibfield  {journal} {\bibinfo  {journal} {J.
  Appl. Phys.}\ }\textbf {\bibinfo {volume} {114}},\ \bibinfo {pages} {194706}
  (\bibinfo {year} {2013})}\BibitemShut {NoStop}%
\bibitem [{\citenamefont {Kitaev}(2001)}]{R30}%
  \BibitemOpen
  \bibfield  {author} {\bibinfo {author} {\bibfnamefont {A.~Y.}\ \bibnamefont
  {Kitaev}},\ }\href@noop {} {\bibfield  {journal} {\bibinfo  {journal} {Phys.
  Usp.}\ }\textbf {\bibinfo {volume} {44}},\ \bibinfo {pages} {131} (\bibinfo
  {year} {2001})}\BibitemShut {NoStop}%
\bibitem [{\citenamefont {Chakraborty}(2007)}]{R31}%
  \BibitemOpen
  \bibfield  {author} {\bibinfo {author} {\bibfnamefont {T.}~\bibnamefont
  {Chakraborty}},\ }\href@noop {} {\emph {\bibinfo {title} {Charge migration in
  DNA}}}\ (\bibinfo  {publisher} {Springer},\ \bibinfo {year}
  {2007})\BibitemShut {NoStop}%
\bibitem [{\citenamefont {Guo}\ and\ \citenamefont
  {f.~Sun}(2012{\natexlab{a}})}]{R32}%
  \BibitemOpen
  \bibfield  {author} {\bibinfo {author} {\bibfnamefont {A.}~\bibnamefont
  {Guo}}\ and\ \bibinfo {author} {\bibfnamefont {Q.}~\bibnamefont {f.~Sun}},\
  }\href@noop {} {\bibfield  {journal} {\bibinfo  {journal} {Phys. Rev. Lett.}\
  }\textbf {\bibinfo {volume} {108}},\ \bibinfo {pages} {218102} (\bibinfo
  {year} {2012}{\natexlab{a}})}\BibitemShut {NoStop}%
\bibitem [{\citenamefont {Guo}\ and\ \citenamefont
  {f.~Sun}(2012{\natexlab{b}})}]{R33}%
  \BibitemOpen
  \bibfield  {author} {\bibinfo {author} {\bibfnamefont {A.}~\bibnamefont
  {Guo}}\ and\ \bibinfo {author} {\bibfnamefont {Q.}~\bibnamefont {f.~Sun}},\
  }\href@noop {} {\bibfield  {journal} {\bibinfo  {journal} {Phys. Rev. B}\
  }\textbf {\bibinfo {volume} {86}},\ \bibinfo {pages} {035424} (\bibinfo
  {year} {2012}{\natexlab{b}})}\BibitemShut {NoStop}%
\bibitem [{\citenamefont {Liu}(2013)}]{R34}%
  \BibitemOpen
  \bibfield  {author} {\bibinfo {author} {\bibfnamefont {W.~V.}\ \bibnamefont
  {Liu}},\ }\href@noop {} {\emph {\bibinfo {title} {Selected Topics in Modern
  Many-Body Theory}}}\ (\bibinfo  {publisher} {Lectures distributed in 2013
  Summer School of Department of Physics at Tsinghua University in Bejing,
  China},\ \bibinfo {year} {2013})\BibitemShut {NoStop}%
\bibitem [{\citenamefont {von Oppen}\ \emph {et~al.}(2017)\citenamefont {von
  Oppen}, \citenamefont {Peng},\ and\ \citenamefont {Pientka}}]{R35}%
  \BibitemOpen
  \bibfield  {author} {\bibinfo {author} {\bibfnamefont {F.}~\bibnamefont {von
  Oppen}}, \bibinfo {author} {\bibfnamefont {Y.}~\bibnamefont {Peng}}, \ and\
  \bibinfo {author} {\bibfnamefont {F.}~\bibnamefont {Pientka}},\ }\href@noop
  {} {\emph {\bibinfo {title} {Topological superconductivity phase in one
  dimension}}}\ (\bibinfo  {publisher} {Oxford University Press},\ \bibinfo
  {year} {2017})\BibitemShut {NoStop}%
\bibitem [{\citenamefont {Gohler}\ \emph {et~al.}(2011)\citenamefont {Gohler},
  \citenamefont {Hamalbeck}, \citenamefont {Markus}, \citenamefont {Hanne},
  \citenamefont {Vager}, \citenamefont {Naaman},\ and\ \citenamefont
  {Zacharias}}]{R36}%
  \BibitemOpen
  \bibfield  {author} {\bibinfo {author} {\bibfnamefont {B.}~\bibnamefont
  {Gohler}}, \bibinfo {author} {\bibfnamefont {V.}~\bibnamefont {Hamalbeck}},
  \bibinfo {author} {\bibfnamefont {T.~Z.}\ \bibnamefont {Markus}}, \bibinfo
  {author} {\bibfnamefont {G.~F.}\ \bibnamefont {Hanne}}, \bibinfo {author}
  {\bibfnamefont {Z.}~\bibnamefont {Vager}}, \bibinfo {author} {\bibfnamefont
  {R.}~\bibnamefont {Naaman}}, \ and\ \bibinfo {author} {\bibfnamefont
  {H.}~\bibnamefont {Zacharias}},\ }\href@noop {} {\bibfield  {journal}
  {\bibinfo  {journal} {Science}\ }\textbf {\bibinfo {volume} {331}},\ \bibinfo
  {pages} {894} (\bibinfo {year} {2011})}\BibitemShut {NoStop}%
\bibitem [{\citenamefont {Schrade}\ \emph {et~al.}(2017)\citenamefont
  {Schrade}, \citenamefont {Thakurathi}, \citenamefont {Reeg}, \citenamefont
  {Hoffman}, \citenamefont {Klinovaga},\ and\ \citenamefont {Loss}}]{R37}%
  \BibitemOpen
  \bibfield  {author} {\bibinfo {author} {\bibfnamefont {C.}~\bibnamefont
  {Schrade}}, \bibinfo {author} {\bibfnamefont {M.}~\bibnamefont {Thakurathi}},
  \bibinfo {author} {\bibfnamefont {C.}~\bibnamefont {Reeg}}, \bibinfo {author}
  {\bibfnamefont {S.}~\bibnamefont {Hoffman}}, \bibinfo {author} {\bibfnamefont
  {J.}~\bibnamefont {Klinovaga}}, \ and\ \bibinfo {author} {\bibfnamefont
  {D.}~\bibnamefont {Loss}},\ }\href@noop {} {\bibfield  {journal} {\bibinfo
  {journal} {Phys. Rev. B}\ }\textbf {\bibinfo {volume} {96}},\ \bibinfo
  {pages} {035306} (\bibinfo {year} {2017})}\BibitemShut {NoStop}%
\end{thebibliography}%

\end{document}